\documentstyle[12pt,aaspp4,amsfonts]{article} 

\slugcomment{\bf resubmitted to AJ: 10/27/99}

\lefthead{DE GRIJS ET AL.}
\righthead{SUPERNOVA REMNANTS IN THE FOSSIL STARBURST IN M82}

\begin{document}

\def\asec {{$\buildrel{\prime\prime}\over .$}}
\def\tims {{$\buildrel{\rm s}\over .$}}

\title{Supernova Remnants in the Fossil Starburst in
M82\footnote{Based on observations with the NASA/ESA {\sl Hubble Space
Telescope}, obtained at the Space Telescope Science Institute, which is
operated by the Association of Universities for Research in Astronomy
(AURA), Inc., under NASA contract NAS 5-26555.}}

\author{Richard de Grijs, Robert W. O'Connell, George D. Becker, and Roger
A. Chevalier}
\affil{Astronomy Department, University of Virginia, PO Box 3818,
Charlottesville, VA 22903-0818; grijs, rwo,
gdb7s, rac5x@virginia.edu} 
\and
\author{John S. Gallagher, {\sc iii}}
\affil{Astronomy Department, University of Wisconsin, 475 North Charter
Street, Madison, WI 53706; jsg@astro.wisc.edu}

\begin{abstract} 
We report the discovery of ten compact H$\alpha$-bright sources in the
post-starburst region northeast of the center of M82, ``M82 B.'' These
objects have H$\alpha$ luminosities and sizes consistent with Type II
supernova remnants (SNRs).  They fall on the same H$\alpha$ surface
brightness-diameter ($\Sigma-D$) relation defined by SNRs in other
nearby star-forming galaxies, with the M82 candidates lying
preferentially at the small diameter end.  These are the first
candidates for optically-visible SNRs in M82 outside the heavily
obscured central starburst within $\sim 250$ pc from the galactic
center.  If these sources are SNRs, they set an upper limit to the end
of the starburst in region ``B2,'' about 500 pc from the galaxy's core,
of $\sim 50$ Myr.  Region ``B1,'' about 1000 pc from the core, lacks
good SNR candidates and is evidently somewhat older.  This suggests star
formation in the galaxy has propagated inward toward the present-day
intense starburst core. 
\end{abstract}


\section{The Fossil Starburst in M82}

M82 is the nearest and best-studied ``starburst'' galaxy (see Telesco
1988 and Rieke et al. 1993 for reviews).  The active starburst has
continued for $\lesssim 20$ Myr at a rate of $\sim 10\,{\rm M}_{\odot}\,
{\rm yr}^{-1}$.  Energy and gas ejection from supernovae, at a rate of
$\sim 0.1$ supernova yr$^{-1}$ (O'Connell \& Mangano 1978, hereafter
OM78; Rieke et al. 1980), drive an H$\alpha$-bright galactic wind along
the minor axis of M82 (e.g., Lynds \& Sandage 1963; McCarthy, Heckman, \&
van Breugel 1987; Shopbell \& Bland-Hawthorn 1998).  The active
starburst is located in the center of the galaxy, and this region has
consequently received intense observational scrutiny.  All of the bright
radio and infrared sources associated with the active starburst are
confined within a radius of $\sim 250$ pc.  Most of this volume is
heavily obscured by dust at optical wavelengths, although visible
structures labeled M82 A, C, E, and F in the nomenclature of OM78,
O'Connell et al. (1995), and Gallagher \& Smith (1999) probably
represent lower extinction parts of the outer starburst.  The optical
energy distributions of regions A and C indicate the presence of massive
ionizing stars with ages of $\sim 5$ Myr (OM78, Marcum \& O'Connell
1996), in agreement with ages based on infrared photometry for the
higher extinction region of the starburst core (e.g., Rieke et al.
1993, Satyapal et al. 1997). 

However, ample evidence exists that this is not the only major starburst
episode to have occurred in M82.  The high surface brightness region M82
B, lying 500--1000 pc NE from the galactic center, has exactly the
properties one would predict for a post-starburst region in which an
older starburst occurred with an amplitude similar to that of the active
burst.  Region B has an A-type absorption-line spectrum dominated by
strong Balmer lines and lacks significant emission lines.  Spectral
synthesis (OM78, Marcum \& O'Connell 1996) implies a sharp main-sequence
cut-off corresponding to an age of $\sim$ 100--200 Myr.  Gallagher \&
Smith (1999) recently obtained an age of $\sim$ 60 Myr for the very
luminous ($M_V \sim -16$) cluster F, located 440 pc SW of the galaxy's
center.  This suggests that active star formation has been propagating
through the disk of M82 during the last $\sim 100$ Myr. 

The present-day starburst core of M82 is well known to contain a large
population of compact supernova remnants (SNRs).  These are obscured at
optical wavelengths by large line-of-sight extinction by dust, but their
structures and evolution have been studied at radio wavelengths (e.g.,
Kronberg 1985, Huang et al.  1994, Muxlow et al.  1994, Allen \&
Kronberg 1998).  Heretofore, no SNRs have been detected at visible
wavelengths in M82.  In this paper we report the discovery of ten SNR
candidates in the ``fossil'' starburst region M82 B, based on
narrow-band H$\alpha$ observations obtained with the {\sl Hubble Space
Telescope (HST)}, and compare their properties to those of known SNRs in
other star-forming galaxies. 

\section{H$\alpha$ Observations and Data Analysis}

To supplement an ongoing broad-band imaging program on super star
clusters in M82 B with the {\sl HST/WFPC2} and {\sl HST/NICMOS} cameras,
we extracted from the {\sl HST} archive H$\alpha$+continuum observations
of the central regions of M82, taken through the F656N narrow-band
filter (March 16, 1997, program GO \#6826; and September 12, 1995, GO
\#5957).  These consisted of pairs of 500s and 300s exposures,
respectively.  We combined the exposures to eliminate cosmic rays using
the IRAF/STSDAS\footnote{The Image Reduction and Analysis Facility
(IRAF) is distributed by the National Optical Astronomy Observatories,
which is operated by the Association of Universities for Research in
Astronomy, Inc., under cooperative agreement with the National Science
Foundation.  STSDAS is the Space Telescope Science Data Analysis System;
its tasks are complementary to the existing IRAF tasks.} tasks {\sc
mkdark} and {\sc cosmicrays}.  From these, we used {\sc
idl}\footnote{The Interactive Data Language ({\sc idl}$^{\rm
\circledR}$) is a registered trademark of Research Systems, Inc.} to
create an H$\alpha$+continuum mosaic of the central regions of M82 and
co-registered this with our own broad-band {\sl HST/WFPC2} images in
{\it V} (F555W) and {\it I} (F814W).  Details of the latter observations
(GO \#7446) will be discussed in de Grijs, O'Connell, \& Gallagher (1999). 

Since we lacked a suitable narrow-band continuum image near the
bandpass of the F656N filter, we created a pseudo-continuum image from
our co-registered (and essentially emission-line-free) {\it V} and
{\it I}-band {\sl WFPC2} images, by linearly interpolating the
continuum fluxes to the mean wavelength of the F656N filter.  We
subtracted the pseudo-continuum image thus constructed from the
H$\alpha$+continuum image to obtain an image containing pure line
emission (the passband of the F656N filter is sufficiently narrow not
to require an [NII] correction).

Figure \ref{fig1.fig} shows the regions of interest here on a
ground-based {\it B}-band image of M82.  Regions A and C are parts of
the starburst core and lie at the base of the minor-axis H$\alpha$
plume, which is not visible in this continuum image (OM78).  The peak of
the 2.2$\micron$ continuum emission from the starburst core, often
called the ``IR nucleus,'' lies 2\arcsec\ west of the center of M82 A
(Dietz et al.  1986, O'Connell et al.  1995).  Region B lies northeast
of A and C, separated from them by the strong central dust lane in the
galaxy.  The box in the image shows the area of M82 B which was imaged
with {\sl HST} in continuum bands by de Grijs et al.  (1999).  Because
there is a gradient of properties within region B, we have divided it
into two subregions: B1 and B2.  The boundary between regions B1 and B2
is RA(J2000) = $9^{\rm h}56^{\rm m}00^{\rm s}$, with B1 lying to the
east of this line. 

We based our flux calibrations on the procedures for {\sl WFPC2} data
recommended by Holtzman et al.  (1995).  In order to check the quality
of the resulting calibration, we compared the total H$\alpha$ fluxes
from the bright regions A and C (OM78) in the {\sl WFPC2} images to
those obtained with 5\asec8 circular aperture, ground-based
spectrophotometry by OM78.  Aperture photometry of our images revealed
the ground-based H$\alpha$ values to be $\sim 20$\% brighter than in the
nominal {\sl WFPC2} calibration.  Considering the various uncertainties,
especially aperture centering errors and our use of an interpolated
continuum for correction of the {\sl HST} emission line data, this is
good agreement.  We will adopt the ground-based zero point for the
H$\alpha$ fluxes discussed in the remainder of this paper. 

In the subsequent analysis, we adopt a distance for M82 of 3.6 Mpc or a
distance modulus of $m-M = 27.8$ mag, based on the Cepheid distance for
M81 obtained by Freedman et al.  (1994). 

\section{H$\alpha$ Emission From the Fossil Starburst Region}

The {\sl HST} pure-H$\alpha$ images as well as ground-based images
(e.g., Lynds \& Sandage 1963, McCarthy et al.  1987) show that there is
relatively little line emission in the disk of M82 at radii larger than
500 pc from the central starburst.  The easily visible H$\alpha$ is
concentrated to the core starburst region and the bright minor-axis
plume extending from it, perpendicular to the galaxy's disk. 

Our continuum-subtracted pure-H$\alpha$ image of region B is shown in
Figure \ref{fig2.fig}a.  There is diffuse H$\alpha$ emission here, but
at much lower levels than in regions A and C and in the bright plume. 
For a quantitative comparison, we integrated the H$\alpha$ flux in our
subtracted image over $20\arcsec \times 20\arcsec$ apertures centered on
regions A, B1, and B2.  Total H$\alpha$ luminosities were $2.6 \times
10^{40}$, $6.7 \times 10^{38}$, and $1.3 \times 10^{39}\, {\rm erg}\,
{\rm s}^{-1}$, respectively.  The H$\alpha$ surface brightness declines
with distance from the galaxy's center. 

Figure \ref{fig2.fig}b shows the same field as Fig.  \ref{fig2.fig}a in
the continuum {\it I\/} band.  A bright diffuse stellar background is
present, and several dozen individual compact star clusters are visible. 
(The clusters are discussed further in de Grijs et al.  1999.) There is
clearly not a one-to-one correspondence between the clusters and the
bright emission line regions.  Many clusters have little or no H$\alpha$
emission, and some H$\alpha$ emission regions have inconspicuous
continuum counterparts. 

Figure \ref{fig2.fig}a shows that region B1 has few compact H$\alpha$
sources.  However, B2, located closer to the active starburst core, is
brighter in H$\alpha$, due to both compact sources and diffuse emission,
although still at a surface brightness $20\times$ lower than in the
active starburst.  We used the {\sc find} routine in {\sc daophot}
(Stetson 1987) to identify compact sources on both the emission line and
continuum images.  We measured net fluxes for these using circular
apertures with radii in the range 3-7.5 pixels (0\asec30 -- 0\asec75),
adjusted depending on the structure and brightness of the local
background.  We obtained H$\alpha$ fluxes and equivalent widths (EWs)
for, respectively, 37 and 50 compact optical continuum sources in B1 and
B2 (cf.  de Grijs et al.  1999), of which only 9 and 16, respectively,
contain significant H$\alpha$ emission.  Positive H$\alpha$ detections
are listed in Table \ref{halpha.tab}. 

Most of the identified H$\alpha$ sources in B2 have H$\alpha$
brightnesses significantly above the norm for the exterior regions of
the galaxy, especially considering the excess extinction in B2
(estimated to have $A_{\rm V} \sim 1.1\pm0.3$ mag higher than B1, de
Grijs et al.  1999).  The brightest object in B2 resolves to a ``string
of pearls'' of discrete sources and is conspicuously located adjacent to
the strong central dust lane which separates region B from the starburst
core.

In addition to H$\alpha$ associated with definite, compact continuum
sources, we found 11 H$\alpha$ compact sources in region B2 with only
faint counterparts in the continuum passbands.  We measured fluxes for
these as above.  Net continuum fluxes were positive, if small, in all
cases, yielding equivalent widths of 50 \AA\ or higher.  These sources
are listed in Table \ref{halpha.tab} with asterisks.  The quoted errors
for the EWs include the estimated statistical uncertainties in the
continuum measures. 

One expects to find two types of compact H$\alpha$ sources in a galaxy
like M82.  H{\sc ii} regions will exist around young star clusters with
ages $\lesssim 10$ Myr by virtue of the presence of ionizing O- or early
B-type stars.  Although very young H{\sc ii} regions can have
substantial H$\alpha$ EWs in the range 100-1000 \AA\ (e.g., Bresolin \&
Kennicutt 1997), there should always be a significant continuum source
associated with these, even if there is considerable extinction in the
vicinity.  Type II supernovae (SNe) can also produce compact H$\alpha$
remnants.  These will often be associated with their parent star
clusters, but in many cases there may be no well-defined compact
continuum source.  Since Type II SNe can occur up to 50 Myr after a star
formation event, the associated cluster may have faded or dynamically
expanded enough to be inconspicuous against the bright background of the
galaxy.  Alternatively, the parent star of the SNR could have been
ejected from the cluster or could have formed initially in the lower
density field.  In fact, earlier studies of resolved starbursts suggest
that 80-90\% of the bright stellar population resides in a diffuse
component outside of compact clusters (e.g., Meurer et al.  1995,
O'Connell et al.  1995). 

The identified M82 B H$\alpha$ sources fall into two wide luminosity
ranges: those with $L ({\rm H}\alpha) < 9 \times 10^{35}\, {\rm erg}\,
{\rm s}^{-1}$ and those with $L({\rm H}\alpha) > 14 \times 10^{35}\,
{\rm erg}\, {\rm s}^{-1}$.  Neither group has the properties expected
for normal H{\sc ii} regions in the disk of a late-type galaxy.  Recent
samples of normal H{\sc ii} regions in spiral and irregular galaxies
have been compiled by, among others, Kennicutt, Edgar \& Hodge (1989),
Bresolin \& Kennicutt (1997), and Youngblood \& Hunter (1999).  The
brightest H$\alpha$ source in M82 B has $L(H\alpha) < 10^{37}\, {\rm
erg}\, {\rm s}^{-1}$.  Although normal galaxies contain many H{\sc ii}
regions with luminosities in this range, they invariably also have much
brighter sources, with luminosities up to 10$^{38-39} {\rm erg}\, {\rm
s}^{-1}$.  The largest M82 B source is smaller than 20 pc in diameter,
whereas typical diameters for normal disk H{\sc ii} regions are 30--100
pc. 

Instead, we believe that the 10 sources in the more luminous group are
good candidates for SNRs.  These are listed in the upper part of Table
\ref{halpha.tab}.  Six of these have only faint counterparts in the
continuum passbands.  Some of the sources (e.g., source \#3 in Fig. 
\ref{fig2.fig}a) show evidence of limb brightening, as might be expected
for older supernova remnants. 

\section{SNR Candidates in M82 B}

H$\alpha$ observations of large numbers of SNRs have recently become
available for a number of nearby galaxies.  In this section, we compare
the properties of our M82 candidate SNRs to those of this larger sample. 
We consider H$\alpha$ luminosities, sizes, and surface brightnesses. 

Figure \ref{fig3.fig} shows that the H$\alpha$ luminosities of the SNR
candidates in M82 B (uncorrected for either foreground or internal
extinction) are consistent with those of the main populations of SNRs in
similar galaxies.  In this figure, we have collected the H$\alpha$
luminosities of SNRs in three late-type galaxies (identified in the
figure legend) as well as for M81, the luminous Sb-type neighbor of M82. 
The M82 values fall at the low end of the plotted range, overlapping
with the SNRs in NGC 300 and M81.  However, a correction for the
obvious, if not well determined, internal extinction in region B
(probably a factor of $\sim 1.5-2.0\times$ at 6500 \AA) would produce
better agreement with the mean of the distribution.  Although we have
chosen these four comparison systems as most relevant to possible SNRs
in M82, the luminosities plotted are typical of the complete sample of
SNRs known in nearby galaxies including M31, M33, M101, and the Large
Magellanic Cloud (Long et al.  1990, Braun \& Walterbos 1993, Yang,
Skillman \& Sramek 1994, Matonick \& Fesen 1997, Gordon et al.  1998). 

A comparison of the physical full-width-half-maxima (FWHM) of our SNR
candidates (see Table \ref{halpha.tab}) with those of the H$\alpha$
emission in the comparison samples reveals that the M82 sources are
relatively small, with sizes averaging 10 pc compared to values of over
50 pc in typical spiral galaxies (see Fig.  \ref{fig4.fig}).  The M82 B
candidates are, however, larger than the bonafide SNR sources in the
central starburst of M82, which have radio continuum diameters in the
range 1--5 pc.  We return to the question of the sizes of our candidates
in \S 5. 

Surface brightness-diameter ($\Sigma-D$) relations in the radio
continuum have formed the basis of many SNR studies; they were first
discussed for Galactic radio SNRs by Woltjer (1972) and Clark \& Caswell
(1976).  At radio wavelengths, the $\Sigma-D$ relations for the Galaxy,
the LMC and other galaxies, including M82 (Berkhuijsen 1986, Huang et
al.  1994, Muxlow et al.  1994), superpose sufficiently well to
construct a composite $\Sigma-D$ relation with relatively small scatter
for all known shell radio SNRs (e.g., Milne, Caswell \& Haynes 1980,
Berkhuijsen 1983, 1986, Green 1984, Case \& Bhattacharya 1998). 

$\Sigma-D$ relations in H$\alpha$ have been discussed before only in the
context of the young SNRs in M33 (Long et al.  1990, Gordon et al. 
1998).  In Figure \ref{fig4.fig}, we plot the H$\alpha$ surface
brightnesses and diameters for the candidate M82 SNRs and a composite
comparison sample.  Surface brightnesses ($\Sigma$) are computed in
units of erg s$^{-1}$ cm$^{-2}$ arcsec$^{-2}$ and are tabulated for our
candidates in Table 1.  The comparison sample includes the same systems
shown in Fig.  \ref{fig3.fig}.  Figure \ref{fig4.fig} shows that a broad
H$\alpha$ $\Sigma - D$ relation exists for SNRs in normal late-type
galaxies.  Although the H$\alpha$ surface brightnesses of the M82
candidates are generally high, and the diameters small, they fall on the
same ($\Sigma-D$) relation as do the normal SNR samples.  This is
additional evidence supporting the interpretation of the M82 candidates
as SNRs. 

High-resolution VLA observations at 8.4 GHz of the M82 central regions
were obtained by Huang et al.  (1994), with a total integration time of
20 minutes, of which the appropriate section covering M82 B was kindly
made available to us by J.  Condon.  None of the H$\alpha$-bright SNR
candidates, with the possible exception of source \#2 (cf.  Fig. 
\ref{fig2.fig}a), appear to show 8.4 GHz radio emission brighter than
$\sim 300 \mu$Jy associated with them.  However, the ``string of
pearls'' of bright, discrete H$\alpha$ sources in the southwestern
corner of our field seems to be an extension of a similar string of
bright 8.4 GHz sources lying behind the strong dust lane just outside
our H$\alpha$ field of view which crosses the galaxy near this position
(see Fig.  \ref{fig1.fig}).  Likewise, most of the bright 8.4 GHz
sources in region B2 are located in areas with higher than average
extinction, where detection of optical or H$\alpha$ counterparts would
be difficult.  Deeper imaging at radio wavelengths may result in
positive detections at lower radio continuum flux limits; due to the
steep spectral index of SNRs, imaging at lower frequencies would be
preferable. 

It is not unusual for SNRs to be H$\alpha$ bright while lacking
significant emission at radio wavelengths.  From a theoretical point of
view, radio and H$\alpha$ emission are due to different physical
processes and are prominent at different evolutionary stages of the
SNR.  Radio emission arises earlier in SNR evolution than does
H$\alpha$ emission.

M31 (Braun \& Walterbos 1993) and M33 (Gordon et al.  1998) are the only
galaxies in which optically-selected SNRs have been studied in detail at
radio wavelengths.  Of the $\sim 150$ optically-selected SNRs and SNR
candidates in M31 and M33, about 40\% are not detected above $\sim 200
\mu$Jy.  Fluxes from similar SNRs in M82 would be expected to be about
10 times fainter than the M31 and M33 cases, so the absence of such
radio emission from the M82 B candidates does not appear to be unusual. 
By contrast, the brightness of the SNRs in the M82 starburst core (cf. 
Huang et al.  1994) may be a product of an abnormally dense interstellar
medium. 

\section{Evolutionary State of the M82 B Supernova Remnants}

Strong H$\alpha$ emission from an SNR ordinarily implies that it is in a
radiative evolutionary state.  H$\alpha$ is observed from nonradiative
remnants, such as Tycho's remnant, but the emission is weak.  Assuming
evolution in a constant density interstellar medium, the radiative state
at a particular radius implies a lower limit for the ambient density. 
Cioffi, McKee, \& Bertschinger (1988) give expressions for the radius
and velocity at which the radiative PDS (pressure-driven snowplow) phase
begins: $R_{\rm PDS} = 5.2 \, E_{51}^{2/7} n_1^{-3/7}$ pc and $v_{\rm
PDS} = 574 \, E_{51}^{1/14} n_1^{1/7}$ km s$^{-1}$, where $E_{51}$ is
the supernova energy in units of $10^{51}$ ergs and $n_1$ is the
hydrogen density in units of 10 cm$^{-3}$.  These expressions assume
solar metallicity, which is appropriate for M82 (OM78).  Given the
smallest diameter of 5.5 pc and taking a standard SN energy, $E_{51}=1$,
the minimum density is 44 cm$^{-3}$.  The shock velocity in such a
remnant is 710 km s$^{-1}$ at the time of shell formation.  The required
density is not unusual for a dense cloud, especially in an active star
formation region. 

An alternative to interstellar interaction is interaction with
pre-supernova mass loss.  The expectation for the SNRs in M82 B is that
their progenitors are at the low mass end of Type II SN progenitors and
thus are approximately B1 to B3 stars when on the main sequence.  These
stars are too cool to have strong winds.  They may have slow, dense
winds as red supergiant stars, but at the radii of interest here, the
density is not expected to be sufficiently high to sustain a radiative
shock front.  Interstellar interaction is the more likely situation. 

Models of radiative shock waves show that the H$\alpha$ intensity is
$\propto n_o v_{\rm sh}^2$, where $n_o$ is the ambient density and
$v_{\rm sh}$ is the shock velocity, if $v_{\rm sh}$ is in the
approximate range 100--200 km s$^{-1}$ and the shocked gas is
pre-ionized (Raymond 1979).  The dependence on $v_{\rm sh}$ is probably
weaker at higher shock velocities and may tend to $\propto n_o v_{\rm
sh}$.  The expansion of a radiative remnant can be approximated by
$v_{\rm sh} \propto n_o^{-0.8} R^{-2.2}$ for a fixed explosion energy
(Chevalier 1974; Cioffi et al.  1988), where $R$ is the radius of the
SNR.  The implied $\Sigma-D$ relation for H$\alpha$ is then
$\Sigma\propto n_o^{-0.6} D^{-4.4}$, perhaps tending to $\Sigma\propto
n_o^{0.2} D^{-2.2}$ at high shock velocities.  For a fixed density, this
relation is considerably steeper than the observed relation shown in
Fig.  \ref{fig4.fig}.  One reason for this may be observational
selection: the observed sample may consist of SNRs for a range of
ambient density which have just become radiative and are therefore
brightest.  They subsequently fade from view and fall below the
detection threshold.  Another reason may be that the expansion takes
places in an inhomogeneous medium, and only a small part of the shell
has become radiative for the apparently small remnants. 

The relation of these possible SNRs to the smaller, nonthermal
radio-emitting remnants in the center of M82 is unclear.  The central
remnants are not observed in optical line emission, so it is unknown
whether they are in a radiative phase of evolution.  These remnants
probably have more massive stars as progenitors and so could have a
complex mass loss environment.  Finally, the higher interstellar
pressure in the center of M82 could lead to different evolutionary
properties. 

\section{Inferred Age of the Post-Starburst Regions}

The presence of SNRs in the post-starburst region of M82 can help to set
limits on its star formation history.  The last SNe in a quenched
starburst region would occur at a time comparable to the longest
lifetime of an SN progenitor after the end of the starburst activity. 
Following Iben \& Laughlin (1989) and Hansen \& Kawaler (1994), the time
$t$ spent between the zero-age main sequence and planetary nebula phase
by an $8 {\rm M}_\odot$ progenitor star, which is generally adopted as a
lower limit for Type II SNe (e.g., Kennicutt 1984), corresponds to $t
\sim 35-55$ Myr.  Type Ia SNe, which involve lower mass stars in binary
systems, can occur much later, but one expects these to be more
uniformly distributed over the galaxy's surface, not concentrated near
regions of recent star formation. 

The radiative lifetimes of the shell SNRs in their later phases are
short in this context.  They are limited by the expansion velocities of
their shells.  Older SNRs will be physically large and have low surface
brightnesses.  Radio observations of well-studied Galactic SNRs indicate
relatively rapid luminosity declines, causing the observable radiative
SNR lifetimes to be $\sim 10^4-10^6$ yr (e.g.,  Braun, Goss \& Lyne 1989;
Shull, Fesen \& Saken 1989; Matthews, Wallace \& Taylor 1998), at which
time they will have attained diameters of order 50--100 pc. 

Therefore, if our candidates are indeed SNRs they suggest an upper limit
to the end of the starburst event in region B2 of $\sim 50$ Myr.  The
absence of SNR candidates in B1 indicates it is older.  The spectral
synthesis dating of region B1 is consistent with this ($\sim$ 100--200
Myr, Marcum \& O'Connell 1996), as are the colors of the {\sl
HST}-detected star clusters (de Grijs et al.  1999).  There is then the
following progression of ages with distance from the center of the
current starburst: B1 ($\ga$ 100 Myr, $r \sim 1000$ pc), B2 ($\la$ 50
Myr, $r \sim 500$ pc), and the present core (active for $\la 20$ Myr, $r
< 250$ pc). 

\section{Summary and Conclusion}

We have studied the H$\alpha$ emission from the post-starburst region
500--1000 pc NE of the center of M82 using {\sl HST/WFPC2} images. 
H$\alpha$ emission in this region in general is far below levels
associated with the current starburst core within $\sim 250$ pc from the
galactic center.  Region B1 has little diffuse emission and few compact
sources.  Region B2, located closer to the core, has more of both.  We
find that the compact sources divide by H$\alpha$ luminosity into two
groups.  The objects in the brighter group are good candidates for Type
II supernova remnants.  Their H$\alpha$ luminosities, surface
brightnesses and sizes are consistent with those of SNRs in other
star-forming galaxies.  We attribute their smaller sizes and higher
surface brightnesses to a relatively higher local interstellar gas
density ($\sim 50\, {\rm cm}^{-3}$) than prevails in normal spiral
disks. 

If these candidates are indeed SNRs they indicate an upper limit to the
end of the starburst event in region B2 of $\sim 50$ Myr and suggest
that intense star-forming activity in M82 has propagated inward toward
the present starburst core during the past 100--200 Myr. 

The evidence that our candidates are actually SNRs is only
circumstantial.  Although their properties seem to compare better to SNR
samples in other galaxies than to those of H{\sc ii} regions, they could
be H{\sc ii} regions which have been affected by the unusual physical
circumstances in M82's disk.  A straightforward test would be to obtain
[S {\sc ii}]/H$\alpha$ emission line ratios for the candidates, since
these are sensitive to the presence of strong shock waves. 

\paragraph{Acknowledgements} This work is based on the undergraduate
senior thesis of George Becker at the University of Virginia.  We are
grateful to Allan Sandage for the loan of the plate reproduced in Fig. 
\ref{fig1.fig} and to Jim Condon for a radio map of M82 B.  We also
acknowledge insightful discussions with Zhi-Yun Li.  This research was
supported by NASA grants NAG 5-3428, NAG 5-6403, and NAG 5-8232.

\newpage

\noindent {\bf FIGURE CAPTIONS}

\figcaption[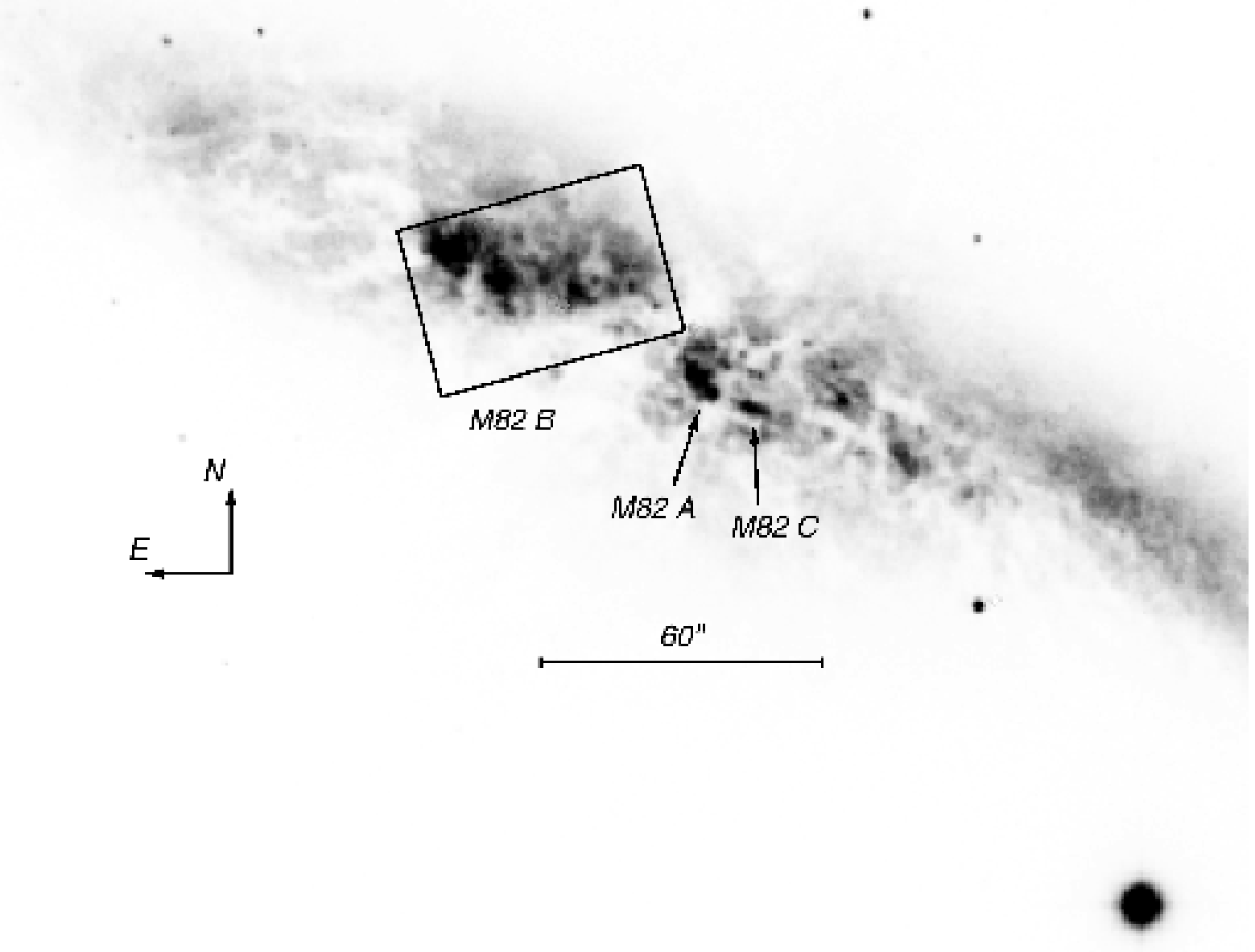]{\label{fig1.fig}A Palomar 5-m plate of M82
taken by Sandage in the {\it B} band (20 minutes, seeing $\la 1
\arcsec$), identifying the regions we discuss.  The image is oriented
north up.  60\arcsec\ corresponds to 1050 pcs.}

\figcaption[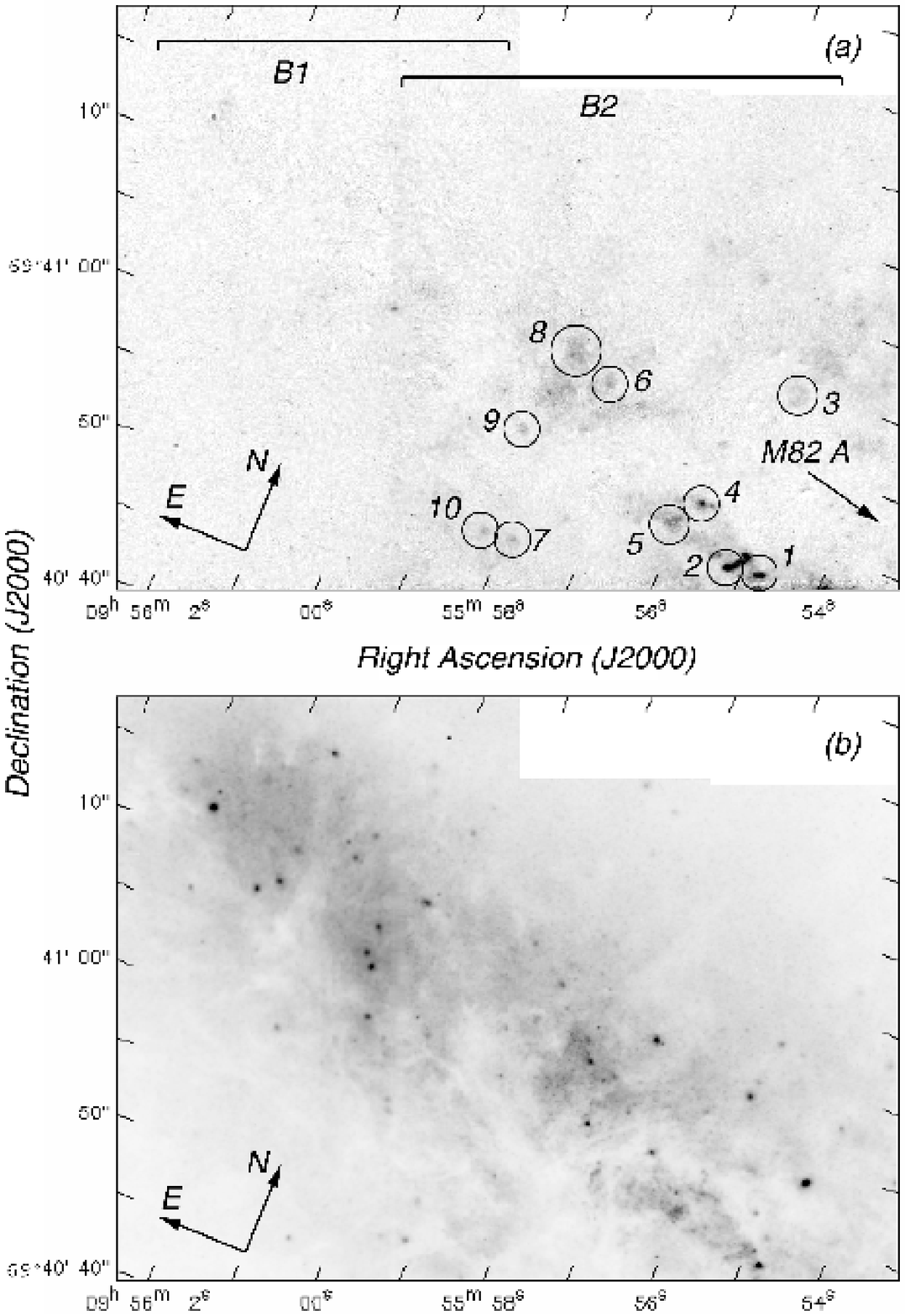]{\label{fig2.fig}{\sl HST/WFPC2} images of
the M82 B region outlined in Fig.  \ref{fig1.fig} (resolution 0\asec1). 
The orientation differs from Fig.  \ref{fig1.fig}.  The 5\arcsec\ bar
corresponds to 88 pc.  {\it (a)} Continuum-subtracted H$\alpha$ image. 
The SNR candidates discussed in this paper are circled.  Numbers
correspond to those in Table \ref{halpha.tab}.  {\it (b)} {\it I}-band
image of same region as in {\it (a)}, showing numerous compact star
clusters, the bright diffuse background, and dust lanes.}

\figcaption[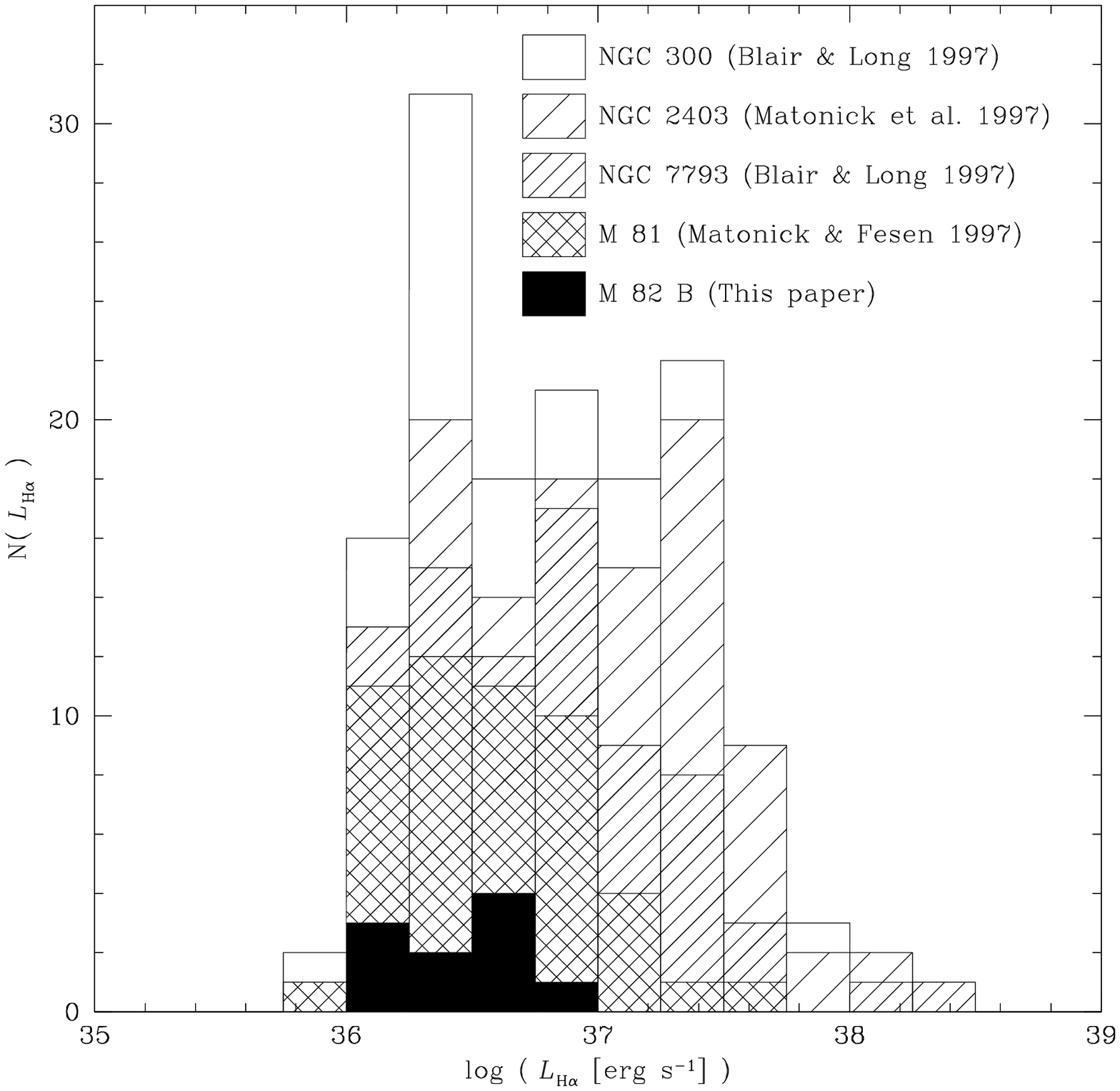]{\label{fig3.fig}Histogram of H$\alpha$
luminosities for SNRs in a sample of normal nearby disk galaxies
compared to those for our M82 SNR candidates.  No extinction corrections
have been applied to the M82 values.}

\figcaption[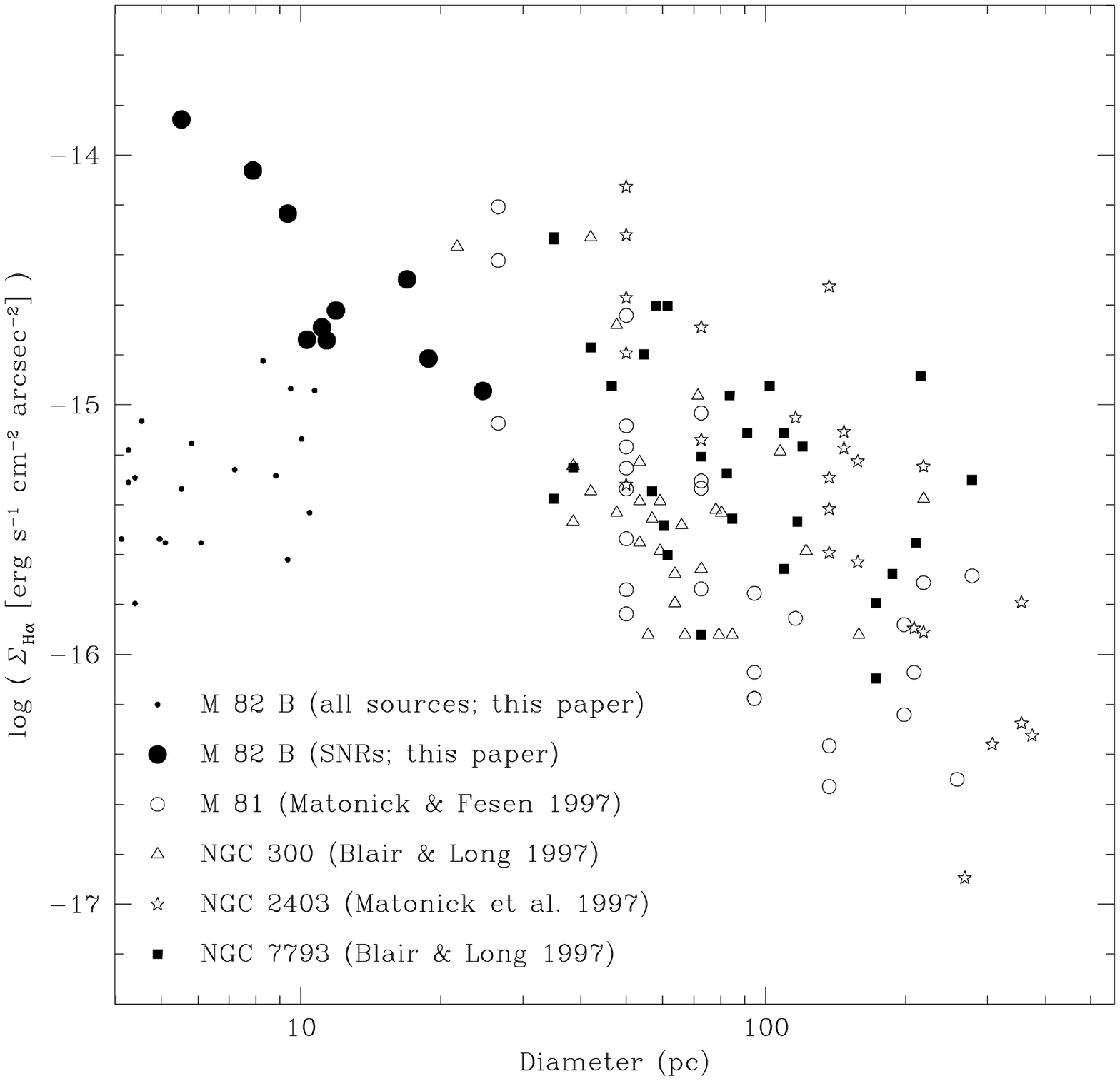]{\label{fig4.fig}Composite H$\alpha$
surface brightness--diameter relation for the SNR candidates in M82 B
and for the normal SNR sample from Fig.  \ref{fig3.fig}.}

\begin{table} 
\scriptsize
\caption[ ]{\label{halpha.tab}Compact H$\alpha$ Sources in M82B}
\begin{center}
\begin{tabular}{crrrcrcrrc} 
\hline 
\hline
No.$^1$  & \multicolumn{1}{c}{R.A.} & \multicolumn{1}{c}{Dec} &
\multicolumn{1}{c}{$L({\rm H}\alpha)$} & $\pm$ &
\multicolumn{1}{c}{$\langle\Sigma_{{\rm H}\alpha}\rangle \times
10^{-16}$} & $\pm$ & \multicolumn{1}{c}{EW} & \multicolumn{1}{c}{$\pm$} &
FWHM$_{{\rm H}\alpha}$ \\
& \multicolumn{2}{c}{(J2000.0)} & \multicolumn{2}{c}{(10$^{35}$ erg
s$^{-1}$)} & \multicolumn{2}{c}{(erg s$^{-1}$ cm$^{-2}$ arcsec$^{-2}$)}
& \multicolumn{2}{c}{(\AA)} & (pc) \\
\hline
\multicolumn{10}{c}{Supernova Remnant Candidates} \\
\phn1 & $09^{\rm h}55^{\rm m}$54\tims71 & $69^\circ40'$55\asec6 & 56.8 & 3.5 &  86.7\phn\phn\phn\phn & 5.3 &  41.5 &   2.6 &  5.3 \\
\phn2 & 55\tims17 & 55\asec3 & 91.0 & 2.2 & 138.9\phn\phn\phn\phn & 3.4 & 354.1 &   8.6 &  3.6 \\
$^*$3 & 55\tims17 & $41'$06\asec5 & 35.6 & 1.8 &  11.4\phn\phn\phn\phn & 0.6 &  50.1 &   4.5 & 18.4\phn \\
\phn4 & 55\tims77 & $40'$58\asec7 & 38.2 & 2.6 &  58.3\phn\phn\phn\phn & 4.0 & 101.7 &   6.8 &  6.4 \\
\phn5 & 55\tims98 & 57\asec0 & 32.5 & 2.2 &  31.8\phn\phn\phn\phn & 2.2 &  26.8 &   1.9 & 12.2\phn \\
$^*$6 & 57\tims48 & $41'$03\asec1 & 23.8 & 1.2 &  23.8\phn\phn\phn\phn & 1.2 &  60.8 &  13.5 &  8.3 \\
$^*$7 & 58\tims03 & $40'$51\asec1 & 14.3 & 0.8 &  18.2\phn\phn\phn\phn & 1.0 & 108.9 &  23.8 &  7.1 \\
$^*$8 & 58\tims03 & $41'$04\asec5 & 19.0 & 1.7 &  15.3\phn\phn\phn\phn & 1.4 &  51.6 &  15.1 & 13.7\phn \\
$^*$9 & 58\tims36 & $40'$58\asec4 & 16.0 & 0.8 &  20.4\phn\phn\phn\phn & 1.0 & 570.1 & 444.5 &  7.7 \\
$^*$10\phn & 58\tims42 & 51\asec2 & 14.2 & 0.7 &  18.1\phn\phn\phn\phn & 0.9 & 214.8 &  65.0 &  7.9 \\
\hline
\multicolumn{10}{c}{Other H$\alpha$ Sources} \\
11 & $09^{\rm h}55^{\rm m}$53\tims95 & $69^\circ41'$11\asec1 & 4.2 & 1.7 &   2.8\phn\phn\phn\phn & 1.1 &   7.4 &   2.9 & $\cdots$ \\
12 & 54\tims33 & 06\asec1 & 1.8 & 1.0 &   2.7\phn\phn\phn\phn & 1.5 &   6.6 &   3.6 &  2.5 \\
13 & 54\tims46 & $40'$56\asec9 & 7.5 & 0.8 &  15.0\phn\phn\phn\phn & 1.6 &  54.1 &   5.6 &  5.6 \\
$^*$14\phn & 54\tims74 & 12\asec5 & 2.2 & 1.0 &   3.7\phn\phn\phn\phn & 1.7 & 114.6 &  75.2 &  7.2 \\
15 & 54\tims77 & 57\asec6 & 1.2 & 1.1 &   1.2\phn\phn\phn\phn & 1.1 &   2.6 &   2.5 & $\cdots$ \\
16 & 55\tims14 & $41'$09\asec8 & 1.6 & 1.0 &   2.4\phn\phn\phn\phn & 1.5 &   6.6 &   4.1 &  2.0 \\
$^*$17\phn & 55\tims85 & 01\asec7 & 7.9 & 0.5 &  17.6\phn\phn\phn\phn & 1.1 & 135.0 &  40.5 & $\cdots$ \\
$^*$18\phn & 56\tims07 & $41'$13\asec3 & 8.9 & 0.8 &  11.4\phn\phn\phn\phn & 1.0 & 307.1 & 126.9 &  7.4 \\
19 & 56\tims48 & $40'$57\asec2 & 0.9 & 0.7 &   2.4\phn\phn\phn\phn & 1.9 &  13.6 &   9.9 &  6.4 \\
$^*$20\phn & 56\tims82 & 14\asec0 & 5.7 & 0.6 &   7.3\phn\phn\phn\phn & 0.8 & 126.5 &  45.0 &  6.9 \\
21 & 57\tims15 & 57\asec6 & 5.2 & 1.2 &   5.1\phn\phn\phn\phn & 1.2 &   9.7 &   2.2 &  2.8 \\
$^*$22\phn & 57\tims45 & $40'$53\asec7 & 5.2 & 0.5 &  11.6\phn\phn\phn\phn & 1.1 & 292.2 & 167.4 &  6.5 \\
23 & 57\tims46 & $41'$00\asec3 & 5.3 & 1.8 &   5.2\phn\phn\phn\phn & 1.8 &   7.5 &   2.5 &  6.0 \\
24 & 57\tims60 & $40'59$\asec7 & 1.8 & 0.6 &   4.9\phn\phn\phn\phn & 1.6 &  38.3 &  13.6 &  2.7 \\
25 & 58\tims01 & $41'$12\asec6 & 1.4 & 0.6 &   2.8\phn\phn\phn\phn & 1.2 &   8.0 &   3.5 &  4.0 \\
26 & 58\tims71 & $40'$58\asec4 & 2.3 & 0.7 &   4.6\phn\phn\phn\phn & 1.4 &  13.5 &   4.2 &  3.6 \\
27 & 59\tims07 & $41'07$\asec1 & 1.6 & 0.6 &   4.3\phn\phn\phn\phn & 1.6 &  19.5 &   7.1 &  1.7 \\
28 & $56'$00\tims02 & 01\asec6 & 4.6 & 0.9 &   7.0\phn\phn\phn\phn & 1.4 &  22.2 &   4.2 &  1.9 \\
29 & 00\tims84 & 08\asec4 & 1.9 & 0.9 &   2.9\phn\phn\phn\phn & 1.4 &   2.0 &   0.9 &  2.6 \\
30 & 00\tims84 & 19\asec8 & 1.1 & 0.6 &   1.6\phn\phn\phn\phn & 0.9 &  94.6 &  52.5 &  2.8 \\
31 & 01\tims25 & 14\asec1 & 4.3 & 1.0 &   6.6\phn\phn\phn\phn & 1.5 &   8.6 &   2.1 &  2.7 \\
32 & 01\tims67 & 07\asec3 & 1.8 & 0.6 &   2.8\phn\phn\phn\phn & 0.9 &   9.3 &   3.1 &  3.3 \\
33 & 03\tims28 & 06\asec7 & 4.6 & 1.0 &   7.0\phn\phn\phn\phn & 1.5 &  18.4 &   3.8 &  3.8 \\
34 & 03\tims33 & 13\asec4 & 2.0 & 1.0 &   5.5\phn\phn\phn\phn & 2.8 &   5.3 &   2.5 &  4.8 \\
35 & 03\tims34 & 12\asec2 & 8.8 & 1.4 &   8.6\phn\phn\phn\phn & 1.4 &   3.4 &   0.6 &  2.9 \\
36 & 03\tims75 & 12\asec8 & 1.4 & 0.7 &   2.9\phn\phn\phn\phn & 1.5 &   7.9 &   3.9 &  3.2 \\
\hline
\end{tabular}
\end{center}
$^1$ -- An asterisk indicates a source with only a faint optical continuum counterpart.

\end{table}

\end{document}